\documentclass[twocolumn,showpacs,preprintnumbers,amsmath,amssymb]{revtex4}

\usepackage[pdftex]{graphicx}
\usepackage{dcolumn}
\usepackage{bm}

\begin{document}


\title{On thermal fluctuations in quantum magnets}

\author{D.A. Tennant$^{1\!,2}$, S. Notbohm$^{1\!,3}$, B. Lake$^{1\!,2}$, A.J.A. James$^{4}$, F.H.L. Essler$^{4}$, H.-J. Mikeska$^{5}$, J. Fielden$^{6}$, P. K\"ogerler$^{6}$, P.C. Canfield$^{6}$,  and M.T.F. Telling$^{7}$}
\affiliation{
 $^1$ Helmholtz-Zentrum Berlin f\"{u}r Materialien und Energie, Glienickerstr. 100, 14109 Berlin, Germany. \\
 $^2$ Institut f\"{u}r Festk\"{o}rperphysik, Technische Universit\"{a}t Berlin, Hardenbergstr. 36,
10623 Berlin, Germany \\
 $^3$ School of Physics and Astronomy, North Haugh, St Andrews, KY15 9SS, U.K.\\ 
 $^4$ Rudolf Peierls Centre for Theoretical Physics, 1 Keble Road, Oxford OX1 3NP, U.K.\\
 $^5$ Department of Theoretical Physics, University of Hannover, Germany \\ 
 $^6$ Ames Laboratory, Iowa State University, IA, U.S.A.\\
 $^7$ ISIS Facility, Rutherford Appleton Laboratory, Chilton, Didcot OX11 OQX, U.K.
 }%

\date{\today}

\begin{abstract}
The effect of thermal fluctuations on the dynamics of a gapped quantum magnet
is studied using inelastic neutron scattering on copper nitrate, a model
material for the one-dimensional (1D) bond alternating Heisenberg chain,
combined with theoretical and numerical analysis. We observe and interpret
the thermally induced central peak due to intraband scattering as well as the
thermal development of an asymmetric continuum of scattering. We relate this
asymmetric line broadening to hard core constraints and quasi-particle
interactions. Our findings are a counter example to recent assertions of
universality of line broadening in 1D systems and are to be considered as a new
paradigm of behaviour, applicable to a broad range of quantum systems. \end{abstract}

\pacs{75.10.Pq, 75.40Gb, 75.50Ee}
\maketitle

The behaviour of quantum systems at finite temperature is of 
great importance for real applications as well as fundamental 
science. In spite of this few experimental studies have hitherto 
been undertaken. The standard paradigm is thermally induced lifetime 
damping of the quasi-particles giving Lorentzian-type 
energy \cite{1, 2, 3} or wave vector \cite{4} broadening. Indeed 
for one-dimensional systems linewidths have been proposed to 
display universality \cite{5,6}. To test the applicability 
of these ideas more widely we studied a generic system of 
quasi-particles subject to hard core interactions in the form 
of the gapped quantum magnet Cu(NO$_3$)$_2$$\cdot2.5$(D$_2$O) by 
use of inelastic neutron scattering. We unambiguously identify 
different aspects of the interplay of quantum and thermal 
fluctuations. The most important finding is the thermal 
development of an {\it asymmetric continuum of scattering}, that 
differs strongly from the Lorentzian broadening familiar 
from conventional theories of thermal decoherence. We propose 
that this paradigm of behaviour is applicable to a broad range 
of low dimensional systems and that this key feature has been 
missed in previous work exhibiting the limitations on the parameter 
regime in which universal behaviour may emerge.

The alternating Heisenberg chain (AHC) of spin-1/2 moments is 
described by the Hamiltonian 
$H=\sum_i J S_{i,1} \cdot S_{i,2} + J' S_{i,2} \cdot S_{i+1,1}$ where 
$J$ and $J'$ are antiferromagnetic exchange constants with $J>J'$ and 
alternation parameter $\alpha$ defined as $J'/J$. The AHC has been an 
important paradigm in quantum magnetism for a long time \cite{7,8,9}. 
Very recently it has attracted the attention of the quantum information 
community as an example for detecting and quantifying entanglement in 
solids \cite{10} as well as encoding and transporting q-bits \cite{11}. 


\begin{figure}[t]
\includegraphics[width=\linewidth]{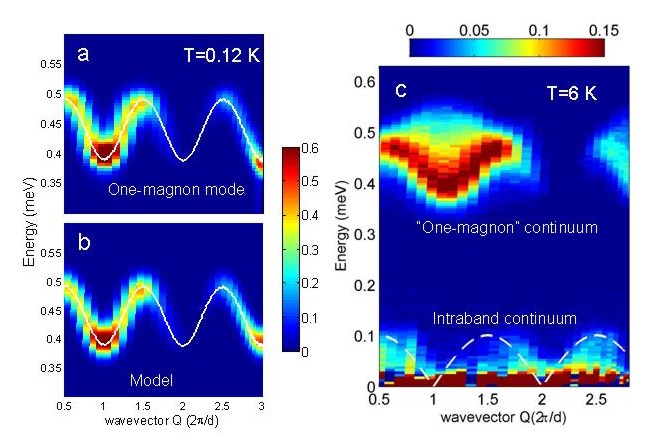}
\caption{\label{fig:fig_1} a-b) One magnon excitations at 0.12K are resolution limited. The dispersion and intensity are explained perfectly by theory, (a) experiment, (b) theory. The wavevector modulation in intensity of the one-magnon band is from an interference term in the scattering amplitude from the two spins within the dimer. c) At 6K the one-magnon scattering has significant asymmetric broadening, and additional scattering appears around E=0meV. This intra-band scattering is found at low temperatures within the superimposed white boundary. }
\end{figure}

In the AHC the dominant antiferromagnetic exchange interaction ($J$) between two neighbouring sites (separated by distance $\rho$ and with chain repeat distance $d$) couples the spins into dimers, whose ground state is a singlet $1/\sqrt{2}\lbrace |\uparrow \downarrow> -|\downarrow \uparrow> \rbrace$. The dimer excited states are a spin-1 triplet with quantum numbers $S^z=1,0,-1$. To first approximation the global ground state of the chain is a product of dimer singlets. However, the inter-dimer exchange ($J'$) admixes a small amount (of order $\sqrt{3}\alpha/8$) of polarised dimer pairs with total spin $s=0$ as ground state fluctuations. The elementary excitations are a triplet of spin-1 states, or magnons. These are momentum $k$ Bloch states obtained by exciting one dimer singlet to a triplet state. The inter-dimer coupling allows the excitation to hop from site-to-site along the chain. To lowest order in $J'$ the one-magnon dispersion is $\omega(k)=J-(J'/2)cos(kd)$ \cite{12}.  

Copper nitrate closely realises the AHC, the Cu$^{2+}$ ions have spin-1/2 moments and the dominant exchange couplings are J=0.443$\pm$0.002 meV and $\alpha$=0.227$\pm$0.005. For these parameters the magnon bandwidth is small compared to the gap. This parameter range is complementary to the one analyzed in recent studies \cite{4,6}. Crucially the smallness of $J'/J$ leads to a clear separation in energy between single magnon excitations and two magnon excitations even at high temperatures.

We measured both the single and multiparticle states for a large (8g) single crystal {\cite{13}} of 98$\%$ deuteration using the OSIRIS neutron spectrometer at the ISIS Facility, UK {\cite{14}}. At very low temperatures compared to the exchange energy $J$, neutrons essentially measure the $S=1$ spectrum by exciting from the ground state. The excitation to the one-magnon modes are easily seen and follow accurately the predicted cosinusoidal behaviour {\cite{14,15}}. The two-quasi-particle states can also be observed. These are sensitive to the interaction between particles and will be used later to quantitatively measure this.

At low temperatures the dynamics of the strongly alternating Heisenberg chain can concomitantly be thought of in terms of a low density gas of hard-core magnons. We follow the thermal evolution of correlations in this quasi-particle gas: The density of thermally excited magnons per-site can be estimated from the zero bandwidth limit to be $n(T)\approx 3\exp(-J/kT)/(1+3\exp(-J/kT))$. Measurements were made at temperatures $T=0.12, 2, 4, 6$ K with densities of $n=6 \cdot 10^{-19}, 0.19, 0.45, 0.56$ respectively (density at infinite temperature is $n(\infty)=0.75$).  With increasing $T$ the following picture emerges: the one-magnon mode develops into a scattering continuum, a new band of scattering emerges around zero energy with the same bandwidth as the one-magnon scattering, and the two-magnon scattering all but disappears, {\it cf} Fig. 1. Below we discuss each of these features in turn.

(i) Temperature dependent lineshape of magnons. The effect of
temperature on the scattering of magnons in copper nitrate was first
investigated by Xu et al. \cite{15} and discussed in the standard
picture of a high temperature (paramagnetic) phase, characterised by
an absence of coherence, and a low temperature phase with
quasi-particles, characterised by exponential decay, giving a Lorentzian
lineshape.

In sharp and remarkable contrast to this picture we find here magnon
lineshapes which are highly asymmetric in energy. Whereas at T=0 the 
one magnon excitations are sharp modes that are resolution limited, comparison 
to Figure 1 shows that the thermal evolution is characterised by the striking
formation of a continuum of scattering weighted towards higher
energies with increasing temperature. Experimental results
are shown in Fig.2 as cuts in energy at wavevectors
$\pi/d$ and $2\pi/d$.

In a phenomenological description of these results we parametrise the
lineshape by fitting it to a continuum with power law singularity at
the lower (or upper) boundary $\omega_{l}$ ($\omega_{u}$) convolved 
with a Lorentzian in energy of width $\Gamma$ representing the damping, 
\begin{eqnarray}
I(Q,\omega)\sim\int d\omega' {{\Gamma_Q{\Theta(\omega'-\omega_l(Q)) \Theta(\omega_u(Q)-\omega')}}\over{2((\omega-\omega')^2+{\Gamma_Q}^2)(\omega' \pm \omega_{l,u})^{\eta}}}.
\end{eqnarray}
This is convolved with the instrumental resolution, modelled by a
Gaussian with variance $\sigma=0.012$ meV. Resulting fits are given as red lines in
fig. 2(a,b), the widths of the Lorentzian and the power law exponent 
are shown in fig. 3(d,e). We argue below that strongly correlated
behaviour from hard core constraints and magnon interactions explain these.


\begin{figure}[tbh]
\includegraphics[width=0.65\linewidth]{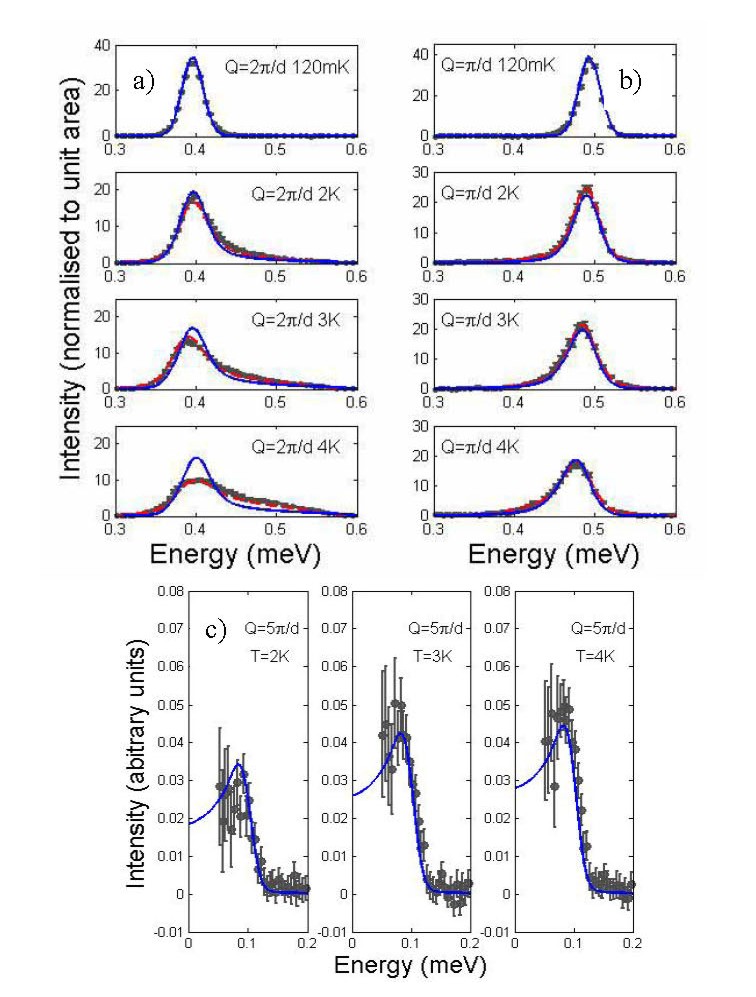}
\caption{\label{fig:fig_4} Temperature dependence of intra-band and one-magnon excitations: The one-magnon scattering at a) $2\pi/d$ and b) $\pi/d$, and c) the intra-band scattering at $5\pi/d$ is given for temperatures 120mK, 2K, 3K and 4K. The red dashed line is a parameterisation using equation 1, and the theory of interacting particles is shown as the solid blue line. Note that for 4K the agreement is less good because the series expansion is only valid at low particle density.}
\end{figure}

(ii) Around zero energy a central peak
in $S(Q,\omega)$ appears due to intraband scattering within the one-magnon band:
Neutrons can scatter via a change in momentum and energy (and possibly spin
quantum number) off a thermally excited magnon. Precisely one
scattering process exists for given wave vector $Q$ and frequency $\omega$ in
a frequency range $\vert \omega \vert < \omega_m(Q)$ (where the maximum
frequency $\omega_m(Q=\pi)$ is the bandwidth). To lowest order we have
$\omega_m(Q) = \alpha J \sin(Q/2)$ and from the density of states a square
root at $\pm \omega_m(Q)$ results. In higher order this singularity becomes
rounded out to a continuum \cite{16}. Results for the lineshape of the central
peak are displayed in Fig. 4(c).  Cuts through the intraband scattering for
$Q=\pi$ for different temperatures are shown in Figure 3(c). Although the
central regime is masked by other incoherent scattering processes from the material, 
the increase towards $\omega_m$ and the drastic decrease beyond $\omega_m$ is clearly seen.


\begin{figure}[tbh]
\includegraphics[width=0.8\linewidth]{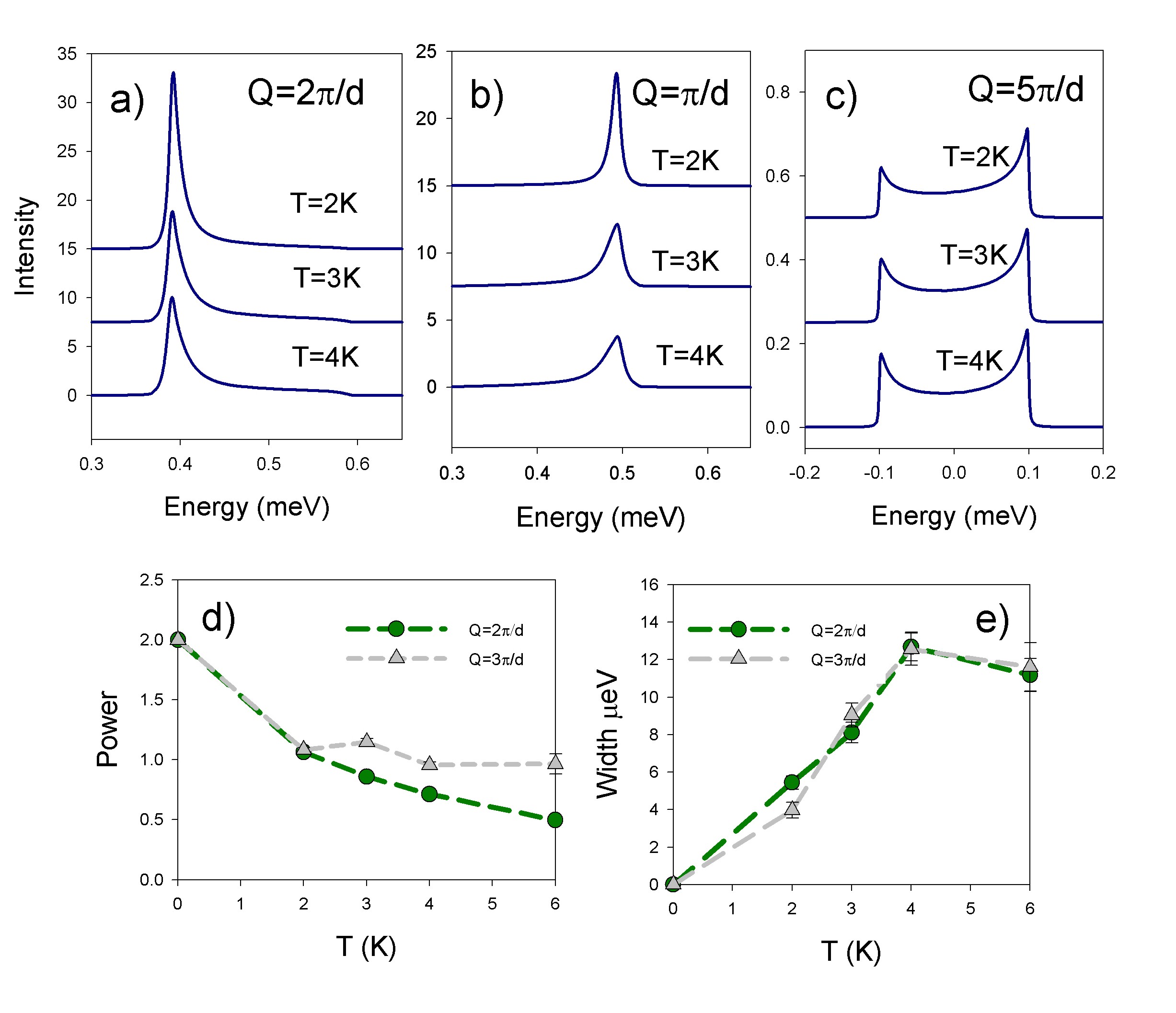}
\caption{\label{fig:fig_2} Asymmetric lineshapes as a function of temperature. The theoretical lineshapes of the one-magnon band at a) $2\pi/d$ and b) $\pi/d$, and of the intra-band at c) $5\pi/d$, calculated using the theory of interactiing particles described in the text. d) and e) give the fitted linewidths and exponents respectively as described by equation (1).}
\end{figure}


\begin{figure}[tbh]
\includegraphics[width=0.75\linewidth]{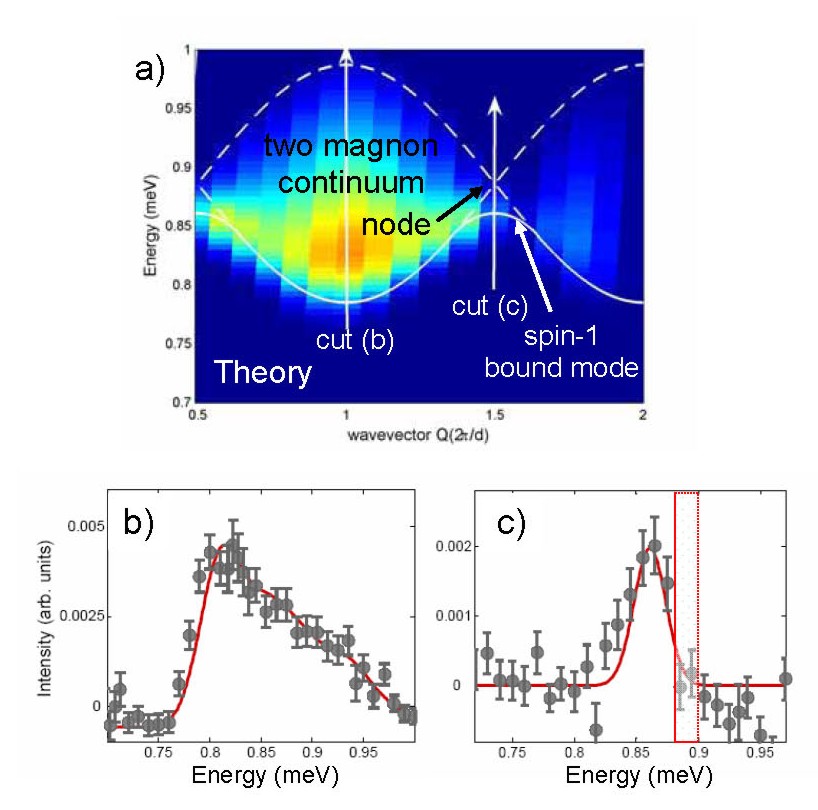}
\caption{\label{fig:fig_0}The two-magnon continuum and bound mode at T=0.12K. a) the superimposed line gives the upper and lower boundaries of the two magnon scattering and the position of the bound mode, lying just below the node. b) is a cut of the scattering intensity at $Q=2\pi/d$ as a function of energy, scattering is observed between the upper and lower two-magnon envelope. c) is a cut at the node position ($Q=3\pi/d$). The node occurs at an energy of 0.88meV, however it is predicted to have very little intensity at this wavevector. Red lines are calculations based on the two particle Schr\"{o}dinger solution [22]. A bound mode is observed at the theoretically predicted lower energy of 0.855meV. }
\end{figure}

The basic square root singularity characterising this scattering process
appears similar to the scattering from excited domain walls in quantum
Heisenberg-Ising chains as predicted by Villain \cite{17} and observed later
in the quasi 1D magnets CsCoCl$_3$ \cite{18,19} and CsCoBr$_3$
\cite{20}. However, this similarity is to large extent formal (cosine spectrum
of the basic excitation) whereas the basic processes behind the scattering are
very different \cite{21}: In the AHC the leading contribution to the intraband
scattering at low temperatures arises from scattering processes between two
well defined single particle states. On the other hand, spin
excitations in the Heisenberg-Ising chain occur in pairs of domain walls and
the leading contribution to the low temperature intraband scattering is due to
scattering on one member of the domain wall pair. Whilst the topological nature of 
these soliton-like excitations restricts it to 1D, the dimer case studied here can serve as
a paradigm for gapped quantum magnets in any number of dimensions.

(iii) Two-magnon scattering and bound states: Neutrons scatter from the weakly admixed polarised dimer pairs composing the ground state fluctuations to the two-magnon states. This scattering is far more sensitive to the composition, interactions and phase factors between particles than the one-magnon signal and so provides key insight. In particular we resolve for the first time the spin-1 bound mode of the AHC which is a direct signature of the two particle interaction, see Figure 4. 

The excitations obey a hard core constraint as each dimer site can be occupied by at most one magnon at a time. In addition to this particles occupying neighbouring sites interact with each other through the inter-dimer exchange ($J'$). This interaction only exists for nearest neighbours and is absent for larger dimer separations. The nearest neighbour potential, $V_s$, is $-J'/2,-J'/4,J'/4$ for pairs with combined total spin $s=0,1,2$ respectively and provides an attraction that has important effects on multi-particle states. There are two types of solution to the two-particle Schr\"{o}dinger equation \cite{22}: The first has the form of a continuum of states, where $k_1$ and $k_2$ are momenta of the magnons along the chain, giving a total momentum $Q=k_1+k_2$ and energy $\hbar\omega=\hbar\omega(k_1)+\hbar\omega(k_2)$. 
The scattering matrix for two particles in a state with total spin $s$ is given by
$S^s_{k_1,k_2}=-{{1+e^{-i(k_1+k_2)d}-2\Delta_s e^{-ik_2 d}}\over{1+e^{-i(k_1+k_2)d}-2\Delta_s e^{-ik_1 d}}}$, where $\Delta_0=0$, $\Delta_1=1/2$, $\Delta_2=-1/2$, and the corresponding phase shift is $\phi^s_{k_1,k_2}=-{i \over 2}\ln(S^s_{k_1,k_2})$. We note that in general the S-matrix differs from the non-interacting hard core value $S_{k_1,k_2}=-1$ but reduces to the latter in the limit of vanishing momenta.

The second solution to the two-particle Schr{\"o}dinger equation has the form of an exponentially decaying bound mode which exists over all $Q$ for $s=0$ and around the nodes in the continuum ($Q\sim\pi/d$) for $s=1$. For $s=2$  the mode occurs at energies above the two-magnon continuum (``anti-bound mode'').
The binding energy and intensity of the bound mode for low temperature agree with the calculations \cite{22} as do the deconfined two particle states, see Figure 4. This experimental observation of the bound mode reveals unambiguously the presence of interparticle interactions.

The decrease in the two-particle scattering with increasing temperature is a direct result of the hard-core constraint. A neutron cannot create a further two magnons on a given pair of dimers if any of the dimers is already occupied by a magnon. The density of adjacent dimer pairs that are both unoccupied falls approximately as $n_{ss}\approx 1/(1+6\exp(-J/kT)+9\exp(-2J/kT))$, i.e. $n_{ss}=1, 0.47, 0.2, 0.14$ respectively for temperatures $0.12, 2, 4, 6 K$ and reaches a value of $n_{ss}(\infty)=0.0625$ at infinite temperature. The intensity of the band is 1 $\%$ of the total scattering at lowest temperatures \cite{22} and with an intensity of only 0.06 $\%$  at high temperature becomes very hard to detect.

To understand further the asymmetric thermal lineshapes for the magnons we
analyzed the $T>0$ dynamical response in the framework of a low density expansion, where the
small parameter is $e^{-J/T}$ \cite{16}. In this regime, lineshapes agree
with those calculated by full diagonalisation of finite chains
\cite{23}. From these diagonalisations one also concludes that the
asymmetry persists beyond the low temperature regime of the analytical
approach, at least up to temperatures of twice the gap.

The asymmetry of the lineshapes results from the combined effect of the hard
core and additional interactions between neighbouring sites, and the density
of states: For hard core particles with no contact potential, i.e. $V=0$, the
quasiparticles are fermionic in the sense that their scattering matrix is
-1. The broadening of the magnon line is
found to be highly asymmetric in this case. The presence of the contact
potential ("stickiness") between quasiparticles {\it further} modifies the
lineshape. This demonstrates that, as expected on general grounds, the
broadening of the line reflects the details of the quasiparticle
interactions. On a qualitative level the broadening of the magnon line can be
understood by considering the joint density of states for transitions between
thermally occupied one magnon states and unoccupied two magnon states
\begin{eqnarray}
\sum_{p,p_1,p_2} n(p)\bar{n}(p_1,p_2) \delta_{q+p,p_1+p_2} \nonumber 
        \delta_{\omega+\omega(p),\omega(p_1)+\omega(p_2)}
\end{eqnarray}
Here $n(p)$ is the thermal occupation number for a one-magnon state with
momentum $p$ and $\bar{n}(p_1,p_2)$ is the probability that the two-magnon state
characterized by momenta $p_1,p_2$ is unoccupied.  For $-\pi/2d < Q < \pi/2d$
and $\pi /2d < Q < 3 \pi/2d$ this function is skewed towards higher and lower
energies respectively, in agreement with experimental (Fig. 2 a,b) and numerical lineshapes. 
The specific form of the lineshape results from the
matrix element and hence by the magnon-magnon interaction, the lines in Figures 
are such calculations.

This behaviour then represents a paradigm for thermal effects in quantum
magnets which differs sharply to the picture of simple lifetime broadening due to thermal decoherence as is observed for example in the 2D Heisenberg antiferromagnet \cite{3}. We expect that such strongly correlated physics is generic to ensembles of interacting hard core particles and should be seen in other quantum gases including other gapped quantum magnets, and in higher dimensions.

In summary, we have performed a comprehensive study of thermal effects in a quantum magnet. Our study shows that temperature does not give rise to simple decoherence but instead promotes the formation of a strongly correlated gas of quasi-partícles. A description of this state has been achieved using non perturbative methods. The findings here present a model of such behavior that should apply to a large range of quantum systems.

We thank Rick Paul (NIST) for measuring the deuteration of the 
sample, S. Pfannenstiel for storage, and K. Damle (TFIR Mumbai) for enlightening discussions. Work at the Ames Laboratory was supported by the Department of Energy, Basic Energy Sciences under Contract No. DE-AC02-07CH11358


\begin{thebibliography}{23}

\bibitem{1} S.P. Bayrakci, T. Keller, K. Habicht and B. Keimer, Science {\bf 312} 1926 (2006).
\bibitem{2} H.M. Ronnow {\it et al.} Phys. Rev. Lett. {\bf 87} 037202 (2001). 
\bibitem{3} T. Huberman, D.A. Tennant, R.A. Cowley, R. Coldea, C.D. Frost, JSTAT P05017 (2008).
\bibitem{4} G.Y. Xu {\it et al.} Science {\bf 317} 1049 (2007).
\bibitem{5} S. Sachdev, "Quantum Phase Transitions", Cambridge University Press (1999).
\bibitem{6} A. Zheludev {\it et al} Phys. Rev. Lett. {\bf 100} 157204 (2008);M. Kenzelmann, R. A. Cowley, W.J.L. Buyers, and D. F. McMorrow, Phys. Rev. B {\bf 63}, 134417 (2001).
\bibitem{7} G.S. Uhrig and H.J. Schulz, Phys. Rev. B{\bf 54} R9624 (1996). 
\bibitem{8} C.J. Hamer, W. Zheng and R.R.P. Singh, Phys. Rev. B{\bf 68}, 214408 (2003).
\bibitem{9} K.P. Schmidt, C. Knetter and G.S. Uhrig, Phys. Rev. B{\bf 69}, 104417 (2004).
\bibitem{10} C. Brukner, V. Vedral and A. Zeilinger, Phys. Rev. A{\bf 73} 012110 (2006).
\bibitem{11} V. Srinivasa, J. Levy and C.S. Hellberg, Phys. Rev. B{\bf 76} 094411 (2007). 
\bibitem{12} T. Barnes, J. Riera and D.A. Tennant, Phys. Rev. B{\bf 59} 11384 (1999).
\bibitem{13} S. Notbohm, "Spin dynamics of quantum spin-ladders and chains" PhD Thesis, U St Andrews, UK (2007).
\bibitem{14} M.T.F. Telling and K.H. Andersen, Phys. Chem. Chem. Phys. {\bf 7} 1255 (2005).
\bibitem{15} G.Y. Xu, C. Broholm, D.H. Reich and M.A. Adams, Phys. Rev. Lett. {\bf 84} 4465 (2000).

\bibitem{16} A.J.A. James, F.H.L. Essler, and R.M. Konik, Phys. Rev. B{\bf 78}, 094411 (2008).
\bibitem{17} J. Villain, Physica B \& C {\bf 79} 1 (1975).
\bibitem{18} J.P. Boucher {\it et al.} Phys. Rev. B{\bf 31} 3015 (1985).
\bibitem{19} H.B. Braun {\it et al.}, Nature Physics {\bf 1} 159 (2005). 
\bibitem{20} S.E. Nagler, W.J.L. Buyers and R.L. Armstrong, Phys. Rev. Lett. {\bf 49} 590 (1982).
\bibitem{21} A.J.A James, W.D. Goetze, and F.H.L. Essler, Phys. Rev. B{\bf 79} 214408 (2009).
\bibitem{22} D.A. Tennant {\it et al.}, Phys. Rev. B{\bf 67} 054414 (2003).
\bibitem{23} H.-J.~Mikeska and C.~Luckmann, Phys. Rev. B{\bf 73} 184426 (2006).


\end{thebibliography}
\end{document}